\documentclass[twocolumn,times,tighten]{aastex631}

\shorttitle{Radio-Loud Exo-Ios }

\begin{document}

%\title{GMRT 200 Ghz Search for Radio-Loud Exo-Ios Orbiting Transiting Hot Jupiters and Hot Saturns:  HD189733b and WASP-49b at 200 GHz searching for radio emission due to planet-moon interaction}

\title{Radio-Loud Exoplanet-Exomoon Survey (RLEES): GMRT Search for Electron Cyclotron Maser Emission}

\author[0000-0002-0554-1151]{Mayank Narang}
\affiliation{Tata Institute of Fundamental Research, Mumbai, India}

\author[0000-0002-1655-0715]{Apurva V. Oza}
\affiliation{Jet Propulsion Laboratory, California Institute of Technology, Pasadena, USA}
\affiliation{Physikalisches Institut, Universit\"{a}t Bern, Bern, Switzerland}

\author[0000-0003-4815-2874]{Kaustubh Hakim}
\affiliation{University of Bern, Center for Space and Habitability, Bern, Switzerland}

\author[0000-0002-3530-304X]{P. Manoj}
\affiliation{Tata Institute of Fundamental Research, Mumbai, India}

\author[0000-0003-0799-969X]{Ravinder K. Banyal}
\affiliation{Indian Institute of Astrophysics, Bangalore, India}

\author[0000-0002-5113-8558]{Daniel P. Thorngren}
\affiliation{Universit\'{e} de Montr\'{e}al, Quebec, Canada}

\begin{abstract}
We conducted the first dedicated search for signatures of exoplanet-exomoon interactions using the Giant Metrewave Radio Telescope (GMRT) as part of the radio-loud exoplanet-exomoon survey (RLEES).  Due to stellar tidal heating, irradiation, and subsequent atmospheric escape, candidate `exo-Io' systems are expected to emit {up to $10^6$ times} more plasma flux than the Jupiter-Io DC circuit.  {This can induce detectable radio emission from the exoplanet-exomoon system.} We analyze three  `exo-Io' candidate stars:  WASP-49, HAT-P 12, and HD 189733.  We perform 12-hour phase-curve observations of WASP-49{b} at 400 MHz during primary $\&$ secondary transit, as well as first $\&$ third quadratures achieving a 3$\sigma$ upper-limit of 0.18 mJy/beam averaged over four days. HAT-P~12 was observed with GMRT at 150 and 325 MHz. We further analyzed the archival data of HD 189733 at 325 MHz. No emission was detected from the three systems. {However, we place strong upper limits on radio flux density.} Given that most exo-Io candidates orbit hot Saturns, we encourage more multiwavelength searches (in particular low frequencies) to span the lower range of exoplanet B-field strengths constrained here.
\end{abstract}
%%%%%%%%%%%%%%%%%%%%%%%%%%%%%%%%%%%%%%%%%%%%%%%%
%AO I removed "will emit" from the abstract. 20 Feb 2022%{may}
% and constrained

%%%%%%%%%%%%%%%%%%%%%%%%%%%%%%%%%%%%%%%%%%%%%%%%
\section{Introduction} \label{Intro}
%%%%%%%%%%%%%%%%%%%%%%%%%%%%%%%%%%%%%%%%%%%%%%%%

Extrasolar satellites (exomoons) have so far eluded ongoing searches due to their small size. Several investigations have exploited transit timing variations leading to the possible identification of giant exomoons \citep[][]{Teachey18, Heller19, Kreidberg19}. Recently, high-resolution spectroscopy has revealed that evaporating exomoons may display alkali metals in hot Jupiter/Saturn atmosphere transit spectra due to their inevitable outgassing due to tidal heating and plasma-driven atmospheric sputtering \citep[][]{Oza19, Gebek20}. These exomoon candidates have been named `exo-Ios' due to their extremely large evaporation rates $\sim 10^{8 \pm 2}$ kg/s (0.2--20 lunar mass/Gyr) capable of catastrophic self-erosion over the often unconstrained age of the star system. 

{One possible method of detecting these elusive exomoons is to search for signals of planet-moon interactions.   
In the solar system, the planet-moon interaction between Jupiter and Io leads to detectable radio emission. The Io-controlled decametric (Io-DAM) emission  \citep{1964Natur.203.1008B} is caused by the motion of Io through Jupiter's magnetic field lines. This motion leads to magnetic field oscillations known as Alfvén waves \citep{Belcher}, which lead to the generation of electric fields parallel to the Jovian magnetic field line \citep{Neubauer80, C97, Saur04, Su09}. As the electrons travel through magnetic field lines, they accelerate and gyrate, leading to radio emission powered by the electron cyclotron maser instability mechanism \citep[ECMI,][]{Wu79, C97}. If a similar mechanism also operates in exomoon-exoplanet systems, then their emission might also be radio-bright. }

{There have been several attempts at detecting star-planet interaction in the  radio domain  \citep[e.g.,][]{Lazio04, Smith09, Hell, Etangs11, Etang13, Turner20, Vedantham20, Prez,Narang21a,2021NatAs...5.1233C}. However, the exoplanet-exomoon interaction has not yet been studied observationally.} In this work, we present the first dedicated survey for studying the planet-moon interaction using the Giant Metrewave Radio Telescope (GMRT) to reveal hidden volcanic exo-Ios (or their exotori counterparts) as well as inform the unknown field strengths of hot Jupiters and hot Saturns.  In Section 2, we describe the targets, followed by the details of the observations and the data reduction process in Section 3.  We describe the results in Section 4. In Section 5, we discuss our findings, followed by a summary in Section 6.

\begin{figure*}
\centering
\includegraphics[width=0.45\linewidth]{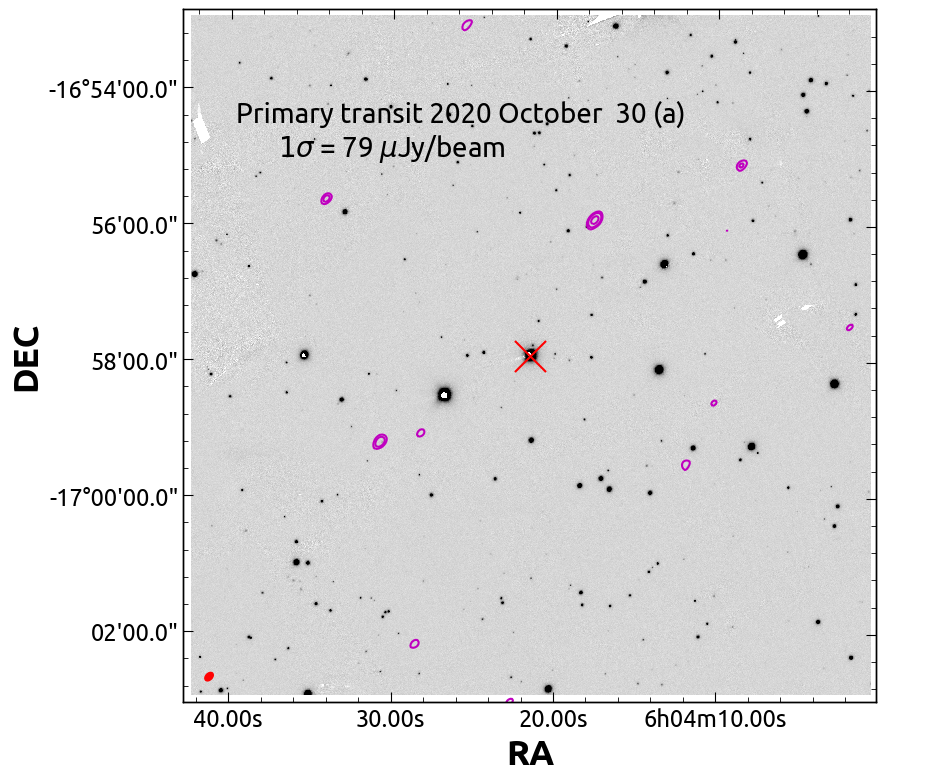}\includegraphics[width=0.45\linewidth]{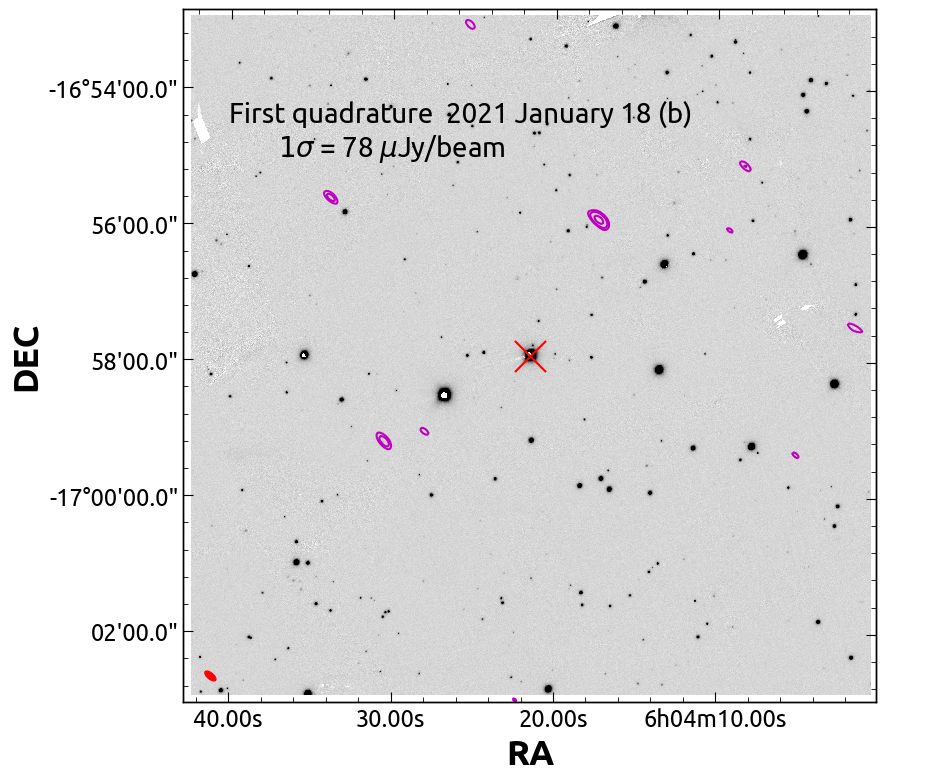}
\includegraphics[width=0.45\linewidth]{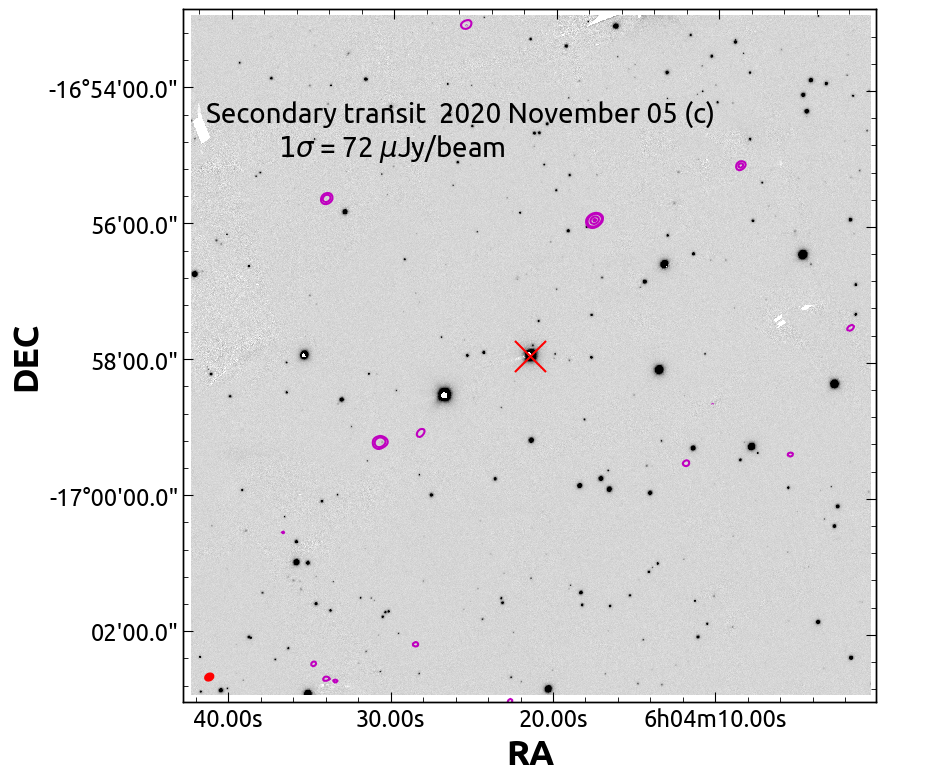}\includegraphics[width=0.45\linewidth]{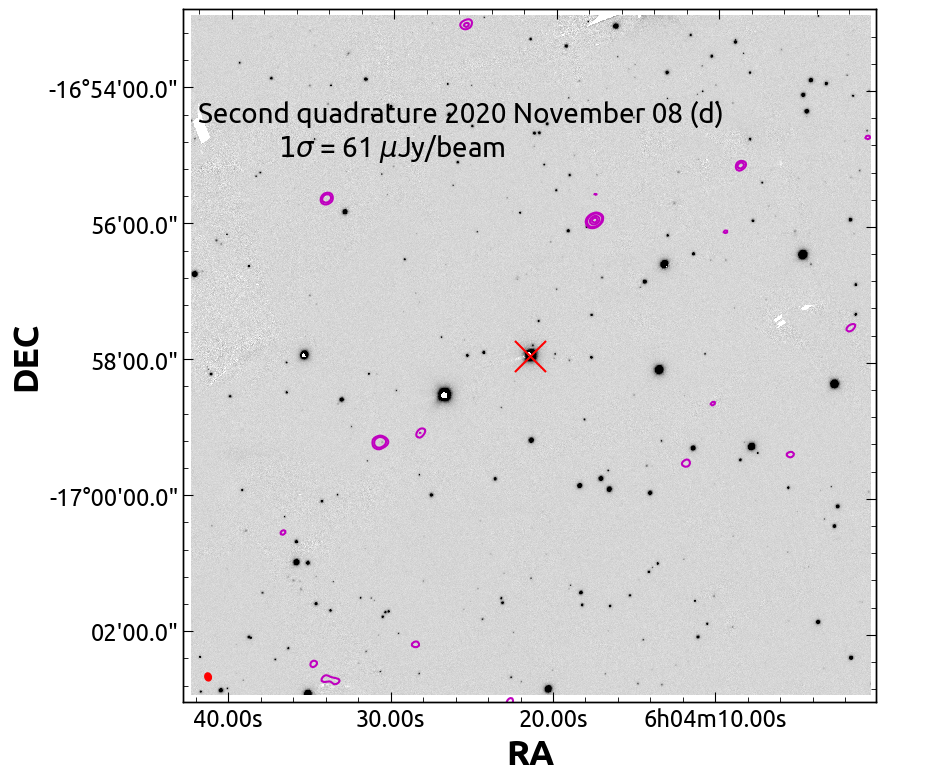}

\caption{The uGMRT image (magenta contours) of the WASP-49 field at 400~MHz  for each of the individual nights of observation overlaid on the PANSTARR g band image. The red cross marks the position of the WASP-49. The contours plotted are  5, 10, 30, and 50~$\times\;\sigma$. The beam is shown as a red ellipse at the bottom left corner. }
\label{fig2}
\end{figure*}

\begin{figure}
\centering
\includegraphics[width=0.9\linewidth]{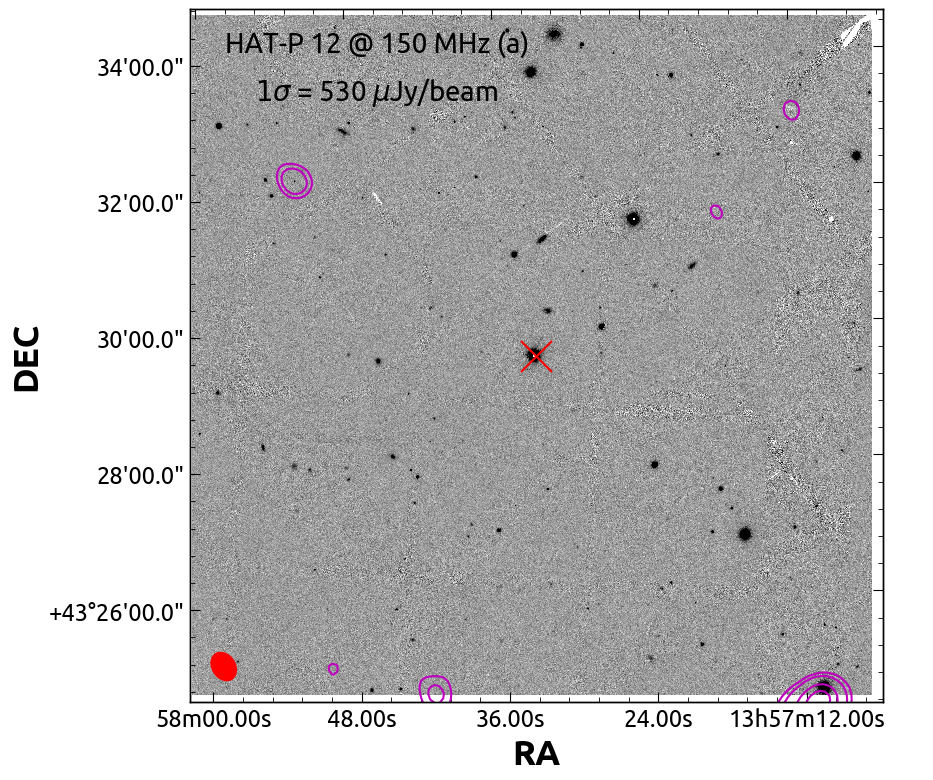}
\includegraphics[width=0.9\linewidth]{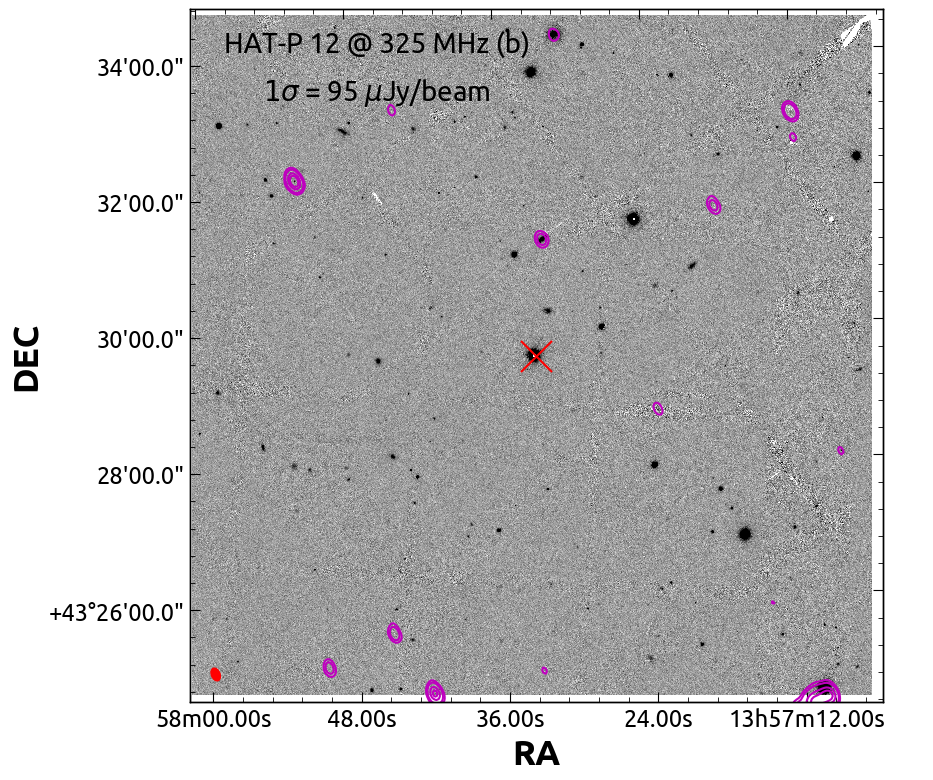}
\caption{The GMRT image (magenta contours) of the HAT-P 12 field at (a) 150~MHz and (b) 325 MHz overlaid on PANSTARR g band image. The red cross marks the position of the HAT-P 12. The contours plotted are  5, 10, 30, and 50~$\times\;\sigma$. The beam is shown as a red ellipse at the bottom left corner. }
\label{fig3}
\end{figure}

\begin{figure}
\centering

\includegraphics[width=0.9\linewidth]{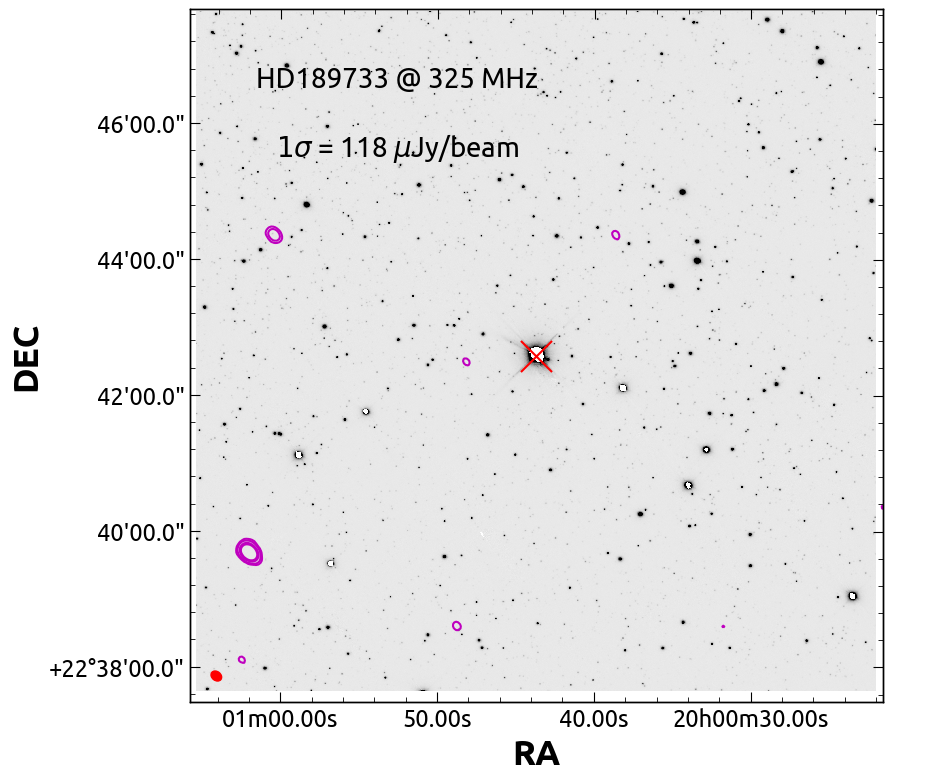}

\caption{The GMRT image (magenta contours) of the HD 189733 field at  325 MHz images overlaid on the PANSTARR g band image. The red cross marks the position of the HD 189733. The contours plotted are  5, 10, 30, and 50~$\times\;\sigma$. The beam is shown as a red ellipse at the bottom left corner. }
\label{fig4}
\end{figure}

%%%%%%%%%%%%%%%%%%%%%%%%%%%%%%%%%%%%
\section{Targets}
%%%%%%%%%%%%%%%%%%%%%%%%%%%%%%%%%%%%

{To select a sample of possible exomoon candidates, we consider planets with alkali exosphere detections in high-resolution spectroscopy \citep[][]{W15, W17, D19}. The main target of our proposal is WASP-49. The WASP-49 system has never been observed at radio wavelengths.  We also retrieve archival GMRT observations of two more exomoon candidates, HD 189733 and HAT-P 12. All three systems are candidate exo-Io systems based on the minimum sodium and potassium column densities implied by high-resolution visible light spectroscopy observations \citep{Oza19}. These potentially evaporating exomoons are well within the tidal stability criterion \citep{Cassidy09}.  Evaporative transmission spectroscopy simulations of two of these systems (WASP-49, HD 189733) demonstrate that an exo-Io or exo-torus scenario is consistent with high-resolution sodium observations at present \citep{Gebek20}. In Table \ref{Table1}, we list the stellar and planetary parameters for WASP-49, HD 189733, and HAT-P 12 systems. }

{The system HD 189733  has been previously observed with GMRT at 150 MHz, 244 MHz, and 614 MHz \citep{Etangs09, Etangs11}. At 150 MHz \cite{Etangs11}  obtained a $3\sigma$ of 2.1 mJy/beam  , while at 244 MHz and 614 MHz \cite{Etangs09} derived a $3\sigma$ upper limit of 2 mJy/beam   and 160 $\mu$Jy/beam   respectively. \cite{Smith09} observed HD 189733 between 304-347 MHz with the Robert C. Byrd Green Bank Telescope of the National Radio Astronomy Observatory (NRAO). The reached and rms sensitivity of 26.7 mJy/beam. No radio observations of HAT-P 12 have been carried out previously. }

\begin{table*}[]
\centering
\resizebox{\textwidth}{!}{%
\begin{tabular}{clccccl}
\hline
Host star & Sp Ty & $M_P$ & $R_P$ & $a_P$ & d & Reference \\
 &  & ($M_J$) & ($R_J$) & (au) & (pc) &  \\ \hline
WASP 49 & G6 V & 0.378 $\pm$ 0.027 & 1.115 $\pm$ 0.047 & 0.0379 $\pm$ 0.001 & $193.73_{-0.52}^{0.68}$ & \cite{Lendl} \\
HAT-P 12 & K4 V & 0.21 $\pm$ 0.01 & $0.959_{-0.021}^{0.029}$ & 0.0384 $\pm$ 0.0003 & 141.75 $\pm$ 0.18 & \cite{Hartman} \\
HD 189733 & K2 V & 1.166 $\pm$ 0.05 & $1.142_{-0.034}^{0.036}$ & 0.031 $\pm$ 0.004 & $19.76_{-0.005}^{0.006}$ & \cite{Addison} \\ \hline
\end{tabular}%
}
\caption{Stellar and planetary parameters of the systems examined in this work. The distance $d$ is from Gaia EDR3/DR3 \cite{B21}.}
\label{Table1}
\end{table*}

\section{Observations and Data Reduction} \label{Observations}

It is unlikely to know a priori the orbital period of exomoons around their parent planets. Moreover, the radio beam, due to planet-moon interaction, may be arbitrarily oriented with respect to the observer. Furthermore, the emission can also be modulated based on the phase of the planet around the star {as argued by \cite{Prez}}. {To maximize the likelihood of detecting the emission, we decided to observe the WASP-49 system at four phases of the planets around the star: the first and second quadrature of the planet WASP-49b, as well as the primary and secondary transit.}  The WASP-49b system was observed for 12 hrs (spread over four observations) with uGMRT in band~3 (250--500 MHz, proposal ID $39\_015$). The center frequency of the receiver was set at 400~MHz, with a bandwidth of 200~MHz. For each observation, the phase center was set at the position of WASP-49b. 

The primary transit of WASP-49b was observed on 2020 October 29$^{th}$. We observed  3C48 as the primary flux density and band-pass calibrator. The flux calibrator was observed twice, once at the beginning of the observation and once at the end of the observation. The phase calibrator used was 0521-207 and was observed in a loop with 30 minutes on the science target and 5 minutes on 0521-207. The secondary transit was observed on 2020, November 5$^{th}$, while the first quadrature (phase 0.25) was observed on  2020, November 8$^{th}$. The observational set-up for these observations was similar to the night of 2020, October 29$^{th}$, with 3C48 as the primary flux density and band-pass calibrator and 0521-207 as the phase calibrator.  The second quadrature (phase 0.75) was observed on 2021, January 17$^{th}$. We used 3C147 as the primary flux density and band-pass calibrator, which was observed at the beginning as well as the end of the observation.  We used  0706-231 as the phase calibrator, which was again observed in a loop of 5 minutes on the phase calibrator and 30 minutes on WASP-49.

We reduce the uGMRT data using the  CASA Pipeline-cum-Toolkit for Upgraded Giant Metrewave Radio Telescope data REduction uGMRT- (CAPTURE) pipeline \citep{CAPTURE}.  We carry out the primary beam correction to correct for the falling sensitivity at the beam edges using the CASA task \textit{wbpbgmrt}\footnote{https://github.com/ruta-k/uGMRTprimarybeam} to produce the final image. 

We further analyze archival GMRT observations of the exoplanet systems HAT-P 12 and HD 189733. The HAT-P 12 field was observed at 150 MHz and 325 MHz with GMRT. At 150 MHz, the system was observed for 11.6 hrs (proposal ID $20\_089$) on 2011, April 28$^{th}$. The phase center of the observation was the HAT-P 12 system. 3C48 was used as the primary flux density, and  1331+305 was used as the phase calibrator. At 325 MHz, the HAT-P 12 field was observed surreptitiously as part of the proposal $22\_051$ on 2012, September 08$^{th}$. The phase center was set to the J1357+43, which is 24.4' away from HAT-P 12. The system was observed for $\sim$7 hrs, with 3C286 being used as the flux calibrator and 1331+305 as the phase calibrator. 

We also retrieve previously unpublished uGMRT observations of the system HD 189733 at 325 MHz. The system was observed for 9.3hrs on 2009, May 26$^{th}$, with the phase center being HD~189733. The primary flux calibrators used were 3C147 and 3C286, while 1924+334 was used as the phase calibrator.  The archival GMRT observations of HAT-P 12  and HD 189733 were reduced using the Source Peeling and Atmospheric Modeling (SPAM) pipeline \citep{Intema14}.  The log of the observations is given in Table \ref{Table2}.

\begin{table*}
\caption{The observation log and rms sensitivity achieved in this work and in literature. }
%AO I wonder if we should put 180 microJy instead of 0.18 mJy in the abstract
\tabcolsep3.0pt $ $
\centering
\begin{tabular}{cccccccccc}
\hline
\\
Target & Frequency & Phase &  Date of   & Bandpass $\&$ Flux density & Phase  & rms  & rms \\
& & &  observation  &    calibrator & calibrator & this work  & literature \\
&[MHz] &-- & --&  -- & -- & [$\mu$Jy/beam] & [$\mu$Jy/beam]  \\ 
\\
\hline
\hline
\\
WASP 49b$^{\dagger}$& 400 & Primary transit & 2020 October 30$^{th}$ & 3C48 & 0521-207 & 79 & -- \\
\\
\hline
\\
WASP 49b$^{\dagger}$ & 400 & First quadrature  & 2021 January 18$^{th}$ &  3C147 & 0706-231 & 78 &-- \\
\\
\hline
\\
WASP 49b$^{\dagger}$ & 400 & Secondary eclipse  & 2020 November 5$^{th}$  & 3C48 & 0521-207 & 72 & --\\
\\
\hline
\\
WASP 49b$^{\dagger}$ & 400 & Second quadrature  & 2020 November 8$^{th}$  & 3C48 & 0521-207
& 61 & --\\
\\
\hline
\hline
\\
HAT-P 12b & 150 & -- & 2011 April 28$^{th}$ & 3C48 & 1331+305 & 530  & --\\
\\
\hline 
\\
HAT-P 12b & 325& -- & 2012 September 08$^{th}$ & 	3C286   & 	1331+305     & 95 & -- \\
\\
\hline
\hline\\
HD 189733b & 325 & -- & 2009 May 26$^{th}$ & 3C147 $\&$ 3C286 & 1924+334 & 118 & 26,667$^a$ \\ \\
\hline\\ \\
%\\
%\hline
%\hline
\end{tabular}\\
$\dagger$P.I: A. Oza 
 \href{https://naps.ncra.tifr.res.in/goa/}{ID 39$\_$015}; a \cite{Smith09}
\label{Table2}
\end{table*}

%%%%%%%%%%%%%%%%%%%%%%%%%%%%%%%%%%%%
\section{Results} \label{Results}
%%%%%%%%%%%%%%%%%%%%%%%%%%%%%%%%%%%%%%%%%%%%%%%%

The WASP-49 field was observed with the uGMRT at 400 MHz for four nights totaling 12 hrs of observation time. The four observations for the WASP-49 field at 400 MHz are shown in Figure \ref{fig2}. No emission was detected for each of the observations. The rms values achieved for each of the four nights are listed in Table \ref{Table2}. Based on these rms values, we put a 3 $\sigma$ upper limit of 0.18 mJy/beam for the emission from this system.  

In Figure \ref{fig3}, we show the archival GMRT observations for the HAT-P 12 system. At 150 MHz, we reached an rms value of 530 $\mu$Jy/beam, and at 325 MHz, we were able to reach an rms value of 95 $\mu$Jy/beam. No radio emission was, however, detected from the system at either of the frequencies. The upper limits of  1.6 mJy/beam at 150 MHz is comparable to some of the deepest limits reached at that frequency for an exoplanet field {\citep[e.g.,][]{Hell, OG18, Narang21a,2022MNRAS.515.2015N}}. The  GMRT observations of  HD 189733  325 MHz are shown in Figure \ref{fig4}. At 325 MHz, the rms value for the HD 189733 field is 95 $\mu$Jy/beam.

\section{Discussion} \label{Discussion}

During our observations, we have produced some of the deepest images of an exoplanet field {\citep[e.g.,][]{ Hell, Etangs11, Etang13, OG18, Prez, Narang21a,2022MNRAS.515.2015N}}. There could be several reasons why no radio emission was detected from these systems. In the following subsection, we discuss some of these possible reasons.

\subsection{Radio-Quiet Exoplanet-Exomoon Emission}

{If the radio emission from exoplanet-exomoon interaction is inherently quiet, in that case, our current instrumentation will not be able to detect it. A major difficulty in our experiment is the sheer distance of the targets; for instance, 2/3 of the candidate exomoon targets we analyzed in this study are located beyond 100 pc; therefore, the flux emitted may be too weak to be detected with uGMRT. Moreover, deeper observations with next-generation radio telescopes are needed to detect radio-quiet exoplanet-exomoon emissions.  }

\subsection{Over estimation of cyclotron frequency and exoplanet magnetic fields}\label{sec:nu_c and Bfields}

The electron cyclotron maser emission is characterized by the maximum cyclotron emission frequency $\nu_{c}$. This frequency for ECMI masers is fundamentally linked to the magnetic field strength $B_0$ of the emitting body  at the radio source location and is given as follows:

\begin{equation}
    \nu_c = 2.8 B_0
    \label{cyclotron}
\end{equation}
{where $B_0$ is in Gauss and $\nu_c$ in MHz.} 

{The observations in this work have been carried out at frequencies in the range of 150--500 MHz. This corresponds to planetary magnetic fields of $\sim$50 --180 G. If the magnetic fields of the exoplanets are lower than these values, then we could have missed the emission. }

To evaluate this possibility, we apply the methods of \cite{Yadav17} to estimate the magnetic fields using evolution modeling \citep{Thorngren2018} to derive the heat flux from the interiors of the planets \citep[see][]{Christensen2009}.  {This gives the mean magnetic field on the dynamo surface as \citep[from][]{Reiners2010}}
\begin{equation}
    B_{rms}^{dyn} \, \mathrm{[kG]} = 4.8\times10^3 (M_P L_P^2)^{1/6}  R_P^{-7/6} .
\end{equation}
{where $M_P$, $L_P$, and  $R_P$ are the mass, luminosity, and radius of the planet (all normalized to solar values).}

However, the dynamo surface is not at the surface of the planet but further in at the liquid-metallic phase transition at approximately 1 Mbar \citep{Yadav17, Chabrier2019}.  To best take this into account, we adapt Eq. 2 of \cite{Yadav17}, which uses a scaling law for the dynamo radius (which was calibrated for planets with $M_P \sim 1 M_J$, to instead use the 1 Mbar radius from our evolution models.  The dipole magnetic field strength at the pole is thus
\begin{equation}
    B_{dipole}^{polar} = \frac{B_{rms}^{dyn}}{\sqrt{2}} \left(\frac{R_{dyn}}{R_P} \right)^3
\end{equation}
{where $R_{dyn}$ is the dynamo radius.}
These equations only consider the dipole portion of the field, which is assumed to be the dominant component.  %As such, the modeled magnetic field outside of the planet is just that of a dipole:
%\begin{equation}
 %   {B}_{r \geq R_P} =
  %      \frac{B_{dipole}^{polar}}{2}
  %      \left(\frac{R_P}{r} \right)^3    
   %     \left[2 \hat{r} \cos(\theta) + \hat{\theta} \sin(\theta)\right].
%\end{equation}

These equations should be seen as rough estimates.  Following \cite{Christensen2009}, we assume that the magnetic field is generated by a dynamo from the release of interior heat (rather than, e.g., rotation).  If this is not the case, then magnetic fields are likely to be weaker; however, observational evidence thus far points toward the strong magnetic field case \citep{Cauley2019}.  Furthermore, we are applying these relations to lower-mass planets (i.e., hot Saturns) than either \cite{Christensen2009} or \cite{Yadav17} were originally considering.  We expect this is still reasonable because the intrinsic temperatures generating the dynamo are comparable, the conductive liquid-metallic region still extends to most of all our planets' radii, and lastly, since we have used modeled dynamo depths rather than the existing scaling relation from \cite{Reiners2010}.

For our most massive exo-Io candidate host HD 189733b, {$M_{P}$ = 1.16 $M_J$}, we find {$B_{dipole}^{polar}$ = 58~G}.  This translates into a maximum cyclotron frequency of 162 MHz for HD 189733b. For  WASP-49b ({$M_{P}$ = 0.38 $M_J$ }) we find {$B_{dipole}^{polar}$ = 85~G}, and for HAT-P-12b {($M_{P}$ = 0.21 $M_J$ }) we find {$B_{dipole}^{polar}$ = 13~G}. These values correspond to a maximum cyclotron frequency of 238 MHz for WASP-49 b and 36.4 MHz for HAT-P-12b.   Hence more observations at lower frequencies are required to comment on the ability and presence of an exo-Io to drive ECMI emission at these systems.

% Dropping my results table here for reference purposes -Daniel
% Planet         B_dynamo    B_pole    Z_planet    rDynamo    mass    radius    flux    age
% HD 209458 b    182.254   120.977     0.16329    1.36101     0.73     1.39    1.06     4
% WASP-69 b       40.1337   15.1822    0.206446   0.901094    0.29     1.11    0.216    2
% WASP-52 b      124.147    66.7866    0.23       1.15938     0.46     1.27    0.647    0.4
% WASP-49 A b    150.745    84.9707    0.368333   1.02921     0.37     1.11    0.794   10
% HD 189733 b    107.913    57.9032    0.151045   1.03068     1.13     1.13    0.445    6
% HAT-P-12 b      35.6036   13.0078    0.318568   0.769531    0.21     0.959   0.19     2.5

\subsection{Time variable and beamed emission}
Radio emissions from planets in our solar system are highly time-variable \citep{Zarka04}. {The decameter emission from Jupiter due to the interaction with Io is highly modulated at scales of milliseconds to days \citep[e.g.,][]{1996GeoRL..23..125Z,2014A&A...568A..53R}}. The radio emission from the exomoon interaction could also be a time variable similar to the variability seen in the Jupiter-Io emission.  {Furthermore, the emission due to the interaction between Io and Jupiter is also highly beamed \citep[e.g.,][]{1998JGR...10326649Q, Zarka04, Ray,2022JGRA..12730160L}. Similar beaming is expected from exomoon-exoplanet interactions.} Long-term monitoring of these systems (WASP-49, HAT-P 12, and HD189733) would be required to rule out the variable or beamed nature of emission. The non-detection could be explained by Earth not being in the cone of emission at the time of observation.

\section{Summary} 

We present the first dedicated search for radio emission at candidate exoplanet-exomoon systems. We analyzed uGMRT/GMRT observations for three systems, WASP-49, HAT-P 12, and HD~189733. We observed WASP-49 in band 3 (300--550 MHz) of uGMRT. We observed the first and second quadrature as well as the primary and secondary transit of the system in order to search for radio emission and its variability from the system. We do not detect any radio emission from the system but place a strong $3\sigma$ upper limit of 0.18 mJy/beam at 400 MHz. We analyzed archival legacy GMRT data for HAT-P 12 and HD 189733. However, no radio emission was detected from both systems.  The HAT-P 12 system was observed at 150 MHz and 325 MHz. At 150 MHz, we obtain a $3\sigma$ upper limit of 1.6 mJy/beam from the HAT-P 12 field; this is one of the deepest images at 150 MHz using GMRT.  A much deeper $3\sigma$ upper limit of 0.21 mJy/beam was reached at 325 MHz for this system.  We further analyzed legacy GMRT observations of the HD 189733 field at 325 MHz. The $3\sigma$ upper limit of 0.36 mJy/beam at 325 MHz is 222 times deeper than the previous observation from \cite{Smith09}.

If an exomoon is present in one of these systems, we detect no radio emission due to time variables and beamed emission, overestimation of the cyclotron frequency, or overestimation of the flux density. The search for radio emission due to planet-moon interaction is an emerging field, and more observations at a lower frequency using LOFAR and GMRT band 2 (120-250 MHz) are perhaps required to detect/rule out the presence of exomoons. Indeed, based on the B-field strengths derived here for hot Saturn hosts \citep{Yadav17}, the majority of exo-Io hosts from \cite{Oza19} would benefit from searches at lower frequencies.  The possibility of evaporating exomoons continues to be tantalizing; however, limited by the characterization of extrasolar gas giant magnetospheres and their interaction with their host stars.

\section*{Acknowledgments}

This work is based on observations made with the Giant Metrewave Radio Telescope, which is operated by the National Centre for Radio Astrophysics of the Tata Institute of Fundamental Research and is located at Khodad, Maharashtra, India. We thank the GMRT staff for efficient support to these observations. We acknowledge support of the Department of Atomic Energy, Government of India, under Project Identification No. RTI4002. Part of this work was conducted at the Jet Propulsion Laboratory, California Institute of Technology, under contract with NASA. KH is supported by the European Research Council via Consolidator Grant ERC-2017-CoG-771620-EXOKLEIN.

\bibliographystyle{aasjournal}

\end{document}